# Direct estimates of nitrogen abundance for Seyfert 2 nuclei

O. L. Dors[1]⋆, M. V. Cardaci[2,3], G. F. Hägele[2,3], M. Valerdi[4], G. S. Ilha[5], C. B. Oliveira[1], R. A. Riffel[6],
S. R. Flury[7,8], K. Z. Arellano-Córdova[9,10], T. Storchi-Bergmann[11], R. Riffel[11], G. C. Almeida[1], I. N. Morais[1]

[1] *Universidade do Vale do Paraíba, Av. Shishima Hifumi, 2911, Cep 12244-000, São José dos Campos, SP, Brazil*
[2] *Facultad de Ciencias Astronómicas y Geofísicas, Universidad Nacional de La Plata, Paseo del Bosque s/n, 1900 La Plata, Argentina*
[3] *Instituto de Astrofísica de La Plata (CONICET-UNLP), La Plata, Avenida Centenario (Paseo del Bosque) S/N, B1900FWA, Argentina*
[4] *Instituto Nacional de Astrofísica, Óptica y Electrónica (INAOE), Luis E. Erro No. 1, Sta. Ma. Tonantzintla, Puebla, C.P. 72840, México.*
[5] *Departamento de Astronomia, Instituto de Astronomia, Geofísica e Ciências Atmosféricas da USP, Cidade Universitária, 05508-900 São Paulo, SP, Brazil*
[6] *Departamento de Física, Centro de Ciências Naturais e Exatas, Universidade Federal de Santa Maria, 97105-900, Santa Maria, RS, Brazil*
[7] *Department of Astronomy, University of Massachusetts Amherst, Amherst, MA 01002, United States*
[8] *NASA FINESST Fellow*
[9] *Institute for Astronomy, University of Edinburgh, Royal Observatory, Edinburgh, EH9 3HJ, UK*
[10] *Department of Astronomy, The University of Texas at Austin, 2515 Speedway, Stop C1400, Austin, TX 78712, USA*
[11] *Departamento de Astronomia, Universidade Federal do Rio Grande do Sul, Av. Bento Gonçalves 9500, Porto Alegre, RS, Brazil*

**ABSTRACT**

We derive the nitrogen and oxygen abundances in the Narrow Line Regions (NLRs) of a sample of 38 local ($z < 0.4$) Seyfert 2 nuclei. For that, we consider narrow optical emission line intensities and direct estimates of the electron temperatures ($T_{\rm e}$-method). We find nitrogen abundances in the range $7.6 < 12+\log(\rm N/H) < 8.6$ (mean value $8.06 \pm 0.22$) or $0.4 < (\rm N/N_{\odot}) < 4.7$, in the metallicity regime $8.3 < 12+\log(\rm O/H) < 9.0$. Our results indicate that the dispersion in N/H abundance for a fixed O/H value in AGNs is in agreement with that for disc H II regions with similar metallicity. We show that Seyfert 2 nuclei follow a similar (N/O)-(O/H) relation to the one followed by star-forming objects. Finally, we find that active galaxies called as "nitrogen-loud" observed at very high redshift ($z > 5$) show N/O values in consonance with those derived for local NLRs. This result indicates that the main star-formation event is completed in the early evolution stages of active galaxies.

## 1 INTRODUCTION

The gas fueling (triggering mechanism) into the Super-Massive Black Hole (SMBH) of Active Galactic Nuclei (AGNs) seems to regulate the nuclear star formation activity. This relation, called as AGN-starburst connection, has been investigated through several studies (e.g. Norman & Scoville 1988; Storchi-Bergmann et al. 2000, 2001; Kauffmann et al. 2007; Diamond-Stanic & Rieke 2012; LaMassa et al. 2013).

The evolution of stars in central regions of galaxies yields a chemical enrichment in the Interstellar Medium (ISM) that can be transported by nuclear spiral arms into the SMBH (e.g. Storchi-Bergmann et al. 2007; Riffel et al. 2016; Brum et al. 2017; Prieto et al. 2019; Schönell et al. 2019). However, other processes also contribute to the AGN chemical content[1], such as:

- accretion of cold/pristine gas on scales of hundreds of parsecs (e.g. Hicks et al. 2013),
- *in situ* star formation in AGN discs (e.g. Huang et al. 2023; Ali-Dib & Lin 2023; Wang et al. 2023),
- transport of low metallicity gas from kpc scales by bars (e.g. Galloway et al. 2015), and
- mergers of galaxies (e.g. Ricci et al. 2017; Ramón-Fox & Aceves 2020).

The processes above can lead to AGNs having a distinct star formation history and, consequently, chemical evolution than those of star-forming nuclei in normal galaxies. In fact, Riffel et al. (2023) showed that the relative fraction of intermediate-age stellar population is higher in AGN hosts when compared to the control sample composite by non-active galaxies (see also Rembold et al. 2017; Mallmann et al. 2018). Also, do Nascimento et al. (2022), by using calibrations between strong emission lines and metallicity (called as strong-line methods), found that Seyfert nuclei have lower metallicities (0.16-0.30 dex) than non-active galaxies. Thus, accurate metallicity ($Z$ or O/H) and elemental abundance (e.g. N/H, S/H) estimates can reveal important information on the fueling processes and star formation that are driving the AGN chemical evolution.

In particular, the abundance of nitrogen (N/H) and its relation with the oxygen (N/O) in the ISM are essential to understanding the nucleosynthesis of stars with distinct masses and the cosmic evolution of galaxies (e.g. Pagel & Edmunds 1981; Henry et al. 2000; Pérez-Montero et al. 2013; Vincenzo & Kobayashi 2018; Johnson et al.

---
⋆ E-mail: olidors@univap.br
[1] For a review see Heckman & Best (2014).

2023). Basically, nitrogen has a secondary origin in massive stars ($M_* \gtrsim 8$ M$_\odot$): it is produced from the conversion of carbon and oxygen during the CNO cycles; and a primary origin in low mass stars: it is produced from the nuclear reactions that involve only hydrogen and helium (e.g. Talbot & Arnett 1974; Clayton 1983; Woosley & Weaver 1995; Arnett 1996; Nomoto et al. 2013).

Concerning the methods to estimate $Z$ and element abundances (e.g. O/H, N/H, S/H) in gaseous nebulae, it is largely accepted that the most reliable method for star-forming regions (SFs, i.e. H II regions, H II galaxies) is the one relied on direct measurements of electron temperatures, called as $T_e$-method[2]. Over decades, the application of the $T_e$-method has made it possible the estimation of nitrogen and oxygen (as other elements) abundances in the gas phase of thousands of SFs in the local universe ($z < 0.4$, e.g. Peimbert & Costero 1969; Searle 1971; Peimbert et al. 1978; Shields & Searle 1978; Pagel et al. 1980; Rayo et al. 1982; Torres-Peimbert et al. 1989; Vilchez & Esteban 1996; van Zee et al. 1998; Kennicutt et al. 2003; Hägele et al. 2006, 2008, 2011, 2012; Bresolin et al. 2009; Gusev et al. 2012; Berg et al. 2013; Zaragoza-Cardiel et al. 2022). In particular, results from the CHAOS project[3] (e.g., Berg et al. 2015; Rogers et al. 2021, 2022) and the Sloan Digital Sky Survey (SDSS, e.g. Izotov et al. 2006; Kojima et al. 2017; Curti et al. 2017) have established an excellent probe of the gas-phase abundances in SFs, confirming the N/O-O/H relation and pointing out the existence of an universal N/O radial gradient in spiral galaxies (Berg et al. 2020). Moreover, recent observations with the James Webb Space Telescope (JWST) have revealed an (unexpected) high ISM chemical enrichment in galaxies at $z > 5$ based on the $T_e$-method (e.g. Arellano-Córdova et al. 2022; Curti et al. 2023; Sanders et al. 2024; Topping et al. 2024).

In contrast, the nitrogen and oxygen abundances in AGNs are ill-defined independently of the redshift. Focusing on AGNs in the local universe, for which optical spectroscopic data are available (e.g. data provided by the MPA/JHU group[4]), the first quantitative nitrogen and oxygen abundance study was carried out by Dors et al. (2017) for a sample of 44 Seyfert 2 nuclei ($z < 0.1$) through a comparison between observational strong emission lines and those predicted by photoionization models. Estimates from the indirect method applied by Dors et al. (2017) are, in principle, less precise than the ones via the $T_e$-method by $\sim 0.1$ dex (see e.g. Denicoló et al. 2002; Hägele et al. 2008). In any case, Dors and collaborators showed that Seyfert 2 nuclei have nitrogen abundances ranging from $\sim 0.3$ to $\sim 7.5$ times the solar value, being the nitrogen of secondary origin and presenting abundance values similar to those of SFs with high metallicity. Additional N/H and O/H estimates of Seyfert 2 nuclei, also by using a strong-line method, were obtained by Pérez-Montero et al. (2019). Finally, Flury & Moran (2020), for the first time and relying on methodology of the $T_e$-method developed for SFs, derived N and O abundances for a sample of narrow-line regions (NLR) of local AGNs, whose data were obtained from the SDSS.

Interestingly, AGNs present a non-explained larger dispersion ($\sim 0.5$ dex) in the N/O-O/H relation in comparison to SFs ($\sim 0.3$ dex, Dors et al. 2017; Pérez-Montero et al. 2019). The origin of this dispersion has a deep impact on the study of the ISM chemical enrichment and neutral gas content in galaxies. Summarly, the scatter in the N/O-O/H relation for SFs can be produced, for instance, by the

following processes (see Schaefer et al. 2020, 2022; Johnson et al. 2023):

• Primary nitrogen, that is originated into intermediate-mass stars [$4 \lesssim (M_*/M_\odot) \lesssim 8$], increases the N/O abundance ratio with time, since these stars release their nitrogen into the ISM much later than the high mass stars enriched the nebula with oxygen (e.g. Pilyugin 1999; Henry et al. 2000; Meynet & Maeder 2002).

• Type WN Wolf-Rayet stars (which have a high mass-loss rate, e.g. Barlow et al. 1981; Crowther 2007) could contribute (if they are present) to a rapid nitrogen enrichment (e.g. Gräfener & Hamann 2008).

• Distinct star formation efficiencies in SFs located in nuclear zones in comparison to those situated along the galactic disks (see e.g. Nicholls et al. 2017).

• Accretion of low metallicity gas (e.g. Hwang et al. 2019; Luo et al. 2021).

Using abundance estimates of a large sample of spiral disk H II regions, Berg et al. (2020) discussed several possible scenarios that could contribute to the N/O–O/H scatter. However, this issue remains as an open question in the gaseous nebulae study. In the case of the N/O-O/H relation for AGNs, it could exist additional sources of scatter acting in these objects. Interestingly, a large scatter is also noted considering the relation between other elements, i.e. Ar/O-O/H (Monteiro & Dors 2021), Ne/O-O/H (Armah et al. 2021), He/H-O/H (Dors et al. 2022), and S/O-O/H (Dors et al. 2023).

Previous nitrogen abundance estimates for a (relatively) large sample of AGNs in the local universe were majority performed through photoionization models (e.g. Dors et al. 2017; Pérez-Montero et al. 2019). These indirect estimates require, for most of the objects, generic photoionization models assuming, for example, simplified spectral energy distributions and geometries, which could differ from the real physical conditions present in the observed objects (e.g. Gruenwald et al. 1997; Ercolano et al. 2003a,b; Monteiro et al. 2004; Wood et al. 2004; Morisset et al. 2005; Ascasibar et al. 2011; Gesicki et al. 2016; Jin et al. 2022; Binette et al. 2023; Zhu et al. 2023), resulting in abundance values less precise than those derived via the $T_e$-method (e.g. Hägele et al. 2008) and introducing doubts about the existence or the magnitude of the N/O-O/H scatter in AGNs.

Dors et al. (2020a) showed, as found for SFs (e.g. Kennicutt et al. 2003), that O/H abundances in AGNs via $T_e$-method are underestimated by $\sim 0.6$ dex (an average value) in comparison those derived through strong-line methods. In a subsequent study, Dors et al. (2020b) showed that this discrepancy is mainly caused by the inappropriate use of relations between temperatures of the low ($T_{\rm low}$) and high ($T_{\rm high}$) ionization gas zones derived for SFs in AGN chemical abundance studies. Using a photoionization model grid, Dors and collaborators derived a new $T_{\rm low}$-$T_{\rm high}$ expression valid for Seyfert 2 nuclei. The use of this new expression yields a better agreement (about 0.2 dex) between O/H AGN estimates via $T_e$-method and strong-line methods. The nitrogen abundance is traditionally derived from the N$^+$/O$^+$ abundance ratio (Peimbert 1967). These ions are located mainly in the low-ionization zone (e.g. Berg et al. 2021), thus, it is required an appropriate estimate of $T_{\rm low}$ in AGNs to derive reliable nitrogen abundances.

In this regard, in the present work, we used optical emission line intensities taken from the SDSS (York et al. 2000) to estimate nitrogen and oxygen abundances via the $T_e$-method for 38 local ($z < 0.2$) Seyfert 2 AGNs. For that, we take advantage of the new methodology of the $T_e$-method for NLRs of AGNs proposed by Dors et al. (2020b). The structure of this paper is as follows. In Section 2 the methodology employed in the abundance estimates (observational data, $T_{\rm e}$-method) is presented. In Sect. 3 the results and their discussion are shown. Conclusions are presented in Sect. 4.

---

[2] For a review of the $T_e$-method see Peimbert et al. (2017) and Pérez-Montero (2017).
[3] https://www.danielleaberg.com/chaos
[4] https://www.sdss4.org/dr17/spectro/galaxy_mpajhu/

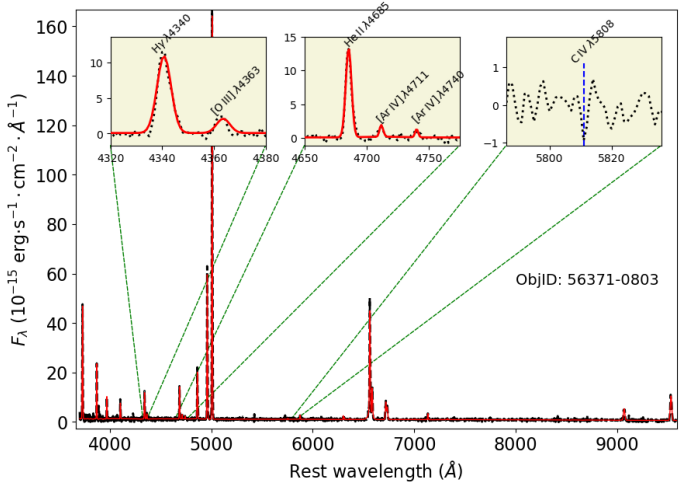

**Figure 1.** Optical spectrum (in black) of one of the Seyfert 2 nuclei in our sample (see Sect. 2.1) taken from the SDSS DR17. The fitting to the emission-line profiles using the IFSCUBE code (Ruschel-Dutra & Dall'Agnol De Oliveira 2020; Ruschel-Dutra et al. 2021) is represented in red color. The measured emission lines and their corresponding wavelengths are indicated. Boxes show zoom-in regions of some weak lines, as indicated.

## 2 METHODOLOGY

### 2.1 Observational data

The observational data used in the present study were taken from the Sloan Digital Sky Survey (SDSS) DR17[5] (Abdurro'uf et al. 2022). The measurements of the emission-line fluxes are obtained using a similar procedure as adopted in Dors et al. (2020a) and summarized in what follows.

To build our sample, we only selected objects for which the [O ii]($\lambda 3726 + \lambda 3729$) (hereafter [O ii]$\lambda 3727$), [O iii]$\lambda 4363$, H$\beta$, He ii$\lambda 4696$, [O iii]$\lambda 5007$, He i$\lambda 5876$, H$\alpha$, [N ii]$\lambda 6584$, and [S ii]$\lambda 6716, \lambda 6731$ emission lines are detected. The signal-to-noise ratio (S/N) for the strong emission lines (all except [O iii]$\lambda 4363$) were required to be (S/N) > 3 and a (S/N) > 2 for the weak auroral line [O iii]$\lambda 4363$. To exclude Seyfert 1 nuclei, which present a prominent secondary ionization/heating source (i. e. shocks, see e.g. Dopita & Sutherland 1996), we follow the criterion originally proposed by Daltabuit & Cox (1972) selecting only objects whose observational permitted emission lines present a Full Width at Half Maximum (FWHM) lower than 1000 km s$^{-1}$. For each spectrum, we carried out the Galactic extinction correction by using the Cardelli et al. (1989) law, assuming the parameterized extinction coefficient $R_V = 3.1$. Thereafter, to obtain the pure nebular spectra, the stellar population continua were subtracted from the observational spectra by using the stellar population synthesis STARLIGHT code (Cid Fernandes et al. 2005). The emission lines were fitted using the publicly available IFSCUBE package (Ruschel-Dutra & Dall'Agnol De Oliveira 2020; Ruschel-Dutra et al. 2021). The fluxes were corrected for extinction, where the observational H$\alpha$/H$\beta$ line ratio was compared with the theoretical value (H$\alpha$/H$\beta$)=2.86 (Hummer & Storey 1987), assuming a temperature of 10 000 K and an electron density of 100 cm$^{-3}$.

---

[5] SDSS DR17 spectroscopic data are available at https://dr17.sdss.org/optical/plate/search

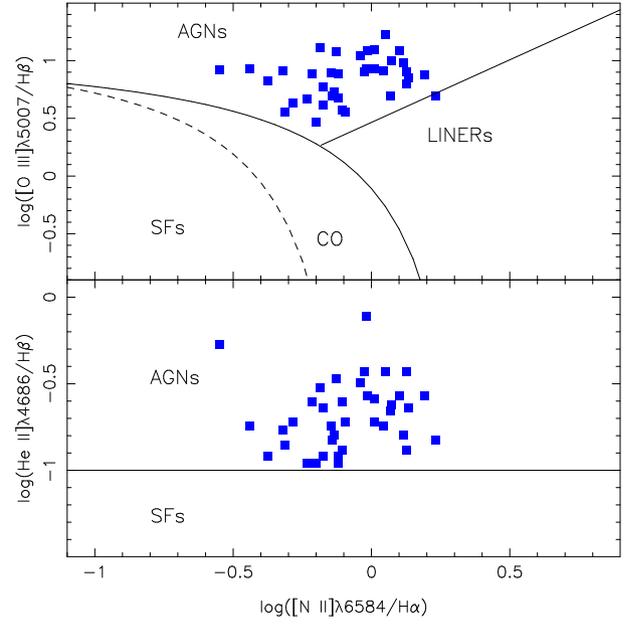

**Figure 2.** Diagnostic diagrams used to separate Star-forming regions (SFs) from Active Galactic Nuclei (AGNs). Blue points represent observational data of our sample of Seyfert 2 nuclei (see Sect. 2.1). Bottom panel: He ii$\lambda 4686$/H$\beta$ versus [N ii]$\lambda 6584$/H$\alpha$ diagram suggested by Shirazi & Brinchmann (2012), where the line represents the criterion by Molina et al. (2021) given by Eq. 1. Top panel: [O iii]/H$\beta$ versus [N ii]/H$\alpha$ (Baldwin et al. 1981) diagram. Solid curve and line represent the criteria proposed by Kewley et al. (2006) to separate SF-AGN and AGN-LINER objects, respectively. The dashed curve represents the empirical criterion proposed by Kauffmann et al. (2003) to separate SF-AGN. Region between the Kewley et al. (2006) and Kauffmann et al. (2003) curves is occupied by Composite Objects (CO).

To select only AGN spectra, we adopted a distinct approach than the one considered in the majority of studies, i.e. we only adopted the He ii$\lambda 4686$/H$\beta$ versus [N ii]$\lambda 6584$/H$\alpha$ diagnostic diagram suggested by Shirazi & Brinchmann (2012), where, according to Molina et al. (2021), objects with

$$\log({\rm He\ II}\lambda 4686/{\rm H}\beta) \gtrsim -1.0 \quad (1)$$

are classified as AGNs, and otherwise as SFs (see also Scholtz et al. 2023). We adopted this criterion instead of the classical BPT diagrams (e.g. [O iii]$\lambda 5007$/H$\beta$ versus [N ii]6584/H$\alpha$, Baldwin et al. 1981) because these diagrams trend to exclude AGNs with low metallicities [$(Z/Z_\odot) \lesssim 0.5$; e.g. Groves et al. 2006; Feltre et al. 2016, 2023; Nakajima & Maiolino 2022; Hirschmann et al. 2023; Harikane et al. 2023; Maiolino et al. 2023; Dors et al. 2024]. In fact, Bykov et al. (2023) used the SRG/eROSITA catalogue[6] of X-ray sources to select 99 AGNs located in dwarf galaxies (galaxies with stellar mass $M_* < 10^{9.5}$ M$_\odot$), which are expected to have low $Z$ (e.g., Izotov & Thuan 2008; Matsuoka et al. 2018; Dors et al. 2019), finding that most of their AGNs are located in the SF zone of the [O iii]$\lambda 5007$/H$\beta$ versus [N ii]6584/H$\alpha$ diagram.

---

[6] https://erosita.mpe.mpg.de/edr/eROSITAObservations/Catalogues/

In Figure 1, the observed spectrum (in black) of an object of our sample, the fitting of the stellar synthesis spectrum (in red) and the weak emission lines (in the boxes) are shown. In Fig. 2, bottom panel, the He II/H$\beta$ versus [N II]/H$\alpha$ diagnostic diagram for our sample is shown. We can see that our data present a wide range (∼ 0.7 dex) of the He II/H$\beta$ emission line intensity ratios, which denotes a large variation of the ionization degree (see e.g. Kehrig et al. 2021 and references therein). On the other hand, the wide range of the [N II]/H$\alpha$ ratio (∼ 0.8 dex) reflects a large variation of metallicity (e.g. Storchi-Bergmann et al. 1994; Denicoló et al. 2002; Groves et al. 2006; Pérez-Montero & Contini 2009; Marino et al. 2013; Carvalho et al. 2020). In Fig. 2, top panel, we can see the position of our sample in the classical [O III]/H$\beta$ versus [N II]/H$\alpha$ diagnostic diagram (Baldwin et al. 1981), where the objects are located in the AGN zone, despite this diagram has not been used as a selection criterion.

The classification of galaxies relying on optical emission lines is under discussion in the literature, mainly due to (relatively) recent data obtained for objects at high redshift (e.g. Sanders et al. 2016, 2023b; Garg et al. 2022). Despite diagrams relying on He II lines do not seem to suffer the uncertainty found in BPT diagrams (e.g. Shirazi & Brinchmann 2012; Dors et al. 2022[7]), these are subject to other caveats. Firstly, the He II$\lambda$4686 emission line is difficult to measure since it is about 4-10 times weaker than H$\beta$ even in nearby galaxies ($z \lesssim 0.1$, e.g. Sartori et al. 2015; Bär et al. 2017; Koss et al. 2017; Wang & Kron 2020; Mayya et al. 2020; Oh et al. 2022; Tozzi et al. 2023). However, the new generation of (extreme) large telescopes and spatial telescopes have shown to circumvent this problem (e.g. Übler et al. 2023). Secondly, the He$^+$ can be ionized, besides the radiation from the accretion disk of AGNs, by:

- certain classes of WR stars, which emit photons with enough energy to ionize nebular He$^+$ (e.g., Kudritzki 2002; Nazé et al. 2003; Sidoli et al. 2006; Crowther 2007; Kehrig et al. 2011; Pérez-Montero et al. 2020; Rickards Vaught et al. 2021), whose ionization potential (IP) is 54.4 eV,
- ultraluminous X-ray (ULX) sources, such as binary stars (e.g. Thuan & Izotov 2005; Schaerer et al. 2019; Simmonds et al. 2021), and
- fast radiative shocks (e.g. Garnett et al. 1991; Plat et al. 2019).

The origin of the He II emission is under discussion in the literature[8] and it is not the goal of the present study to investigate this issue. We assume that, for our sample, He$^+$ and other elements the main ionization mechanism is the hard radiation from the accretion disks of AGNs. This assumption is based on what follows. First, two important broad features in the optical spectra can reveal the presence of WR stars (see López-Sánchez & Esteban 2008 and references therein): the so-called blue WR bump (between 4650–4690 Å) and the red WR bump (formed by the C IV$\lambda$5808 emission line). We did not detect in the spectra of our AGN sample any presence of such WR bumps. In Fig. 1, right-top box, the wavelength region showing the absence of the C IV$\lambda$5808 line in a spectrum of our sample is indicated. Probably, deeper spectroscopic observations (e.g. Alloin et al. 1992; Mazzalay et al. 2010; Revalski et al. 2018b,a, 2022) could reveal the presence of WR features. In any case, we can assume that this class of stars does not drive the He II emission in the galaxies considered here. Second, concerning ULX ionization, Díaz Tello et al. (2017), relying on observational data obtained by the Chandra

---

[7] Dors et al. (2022) suggested the [O III]$\lambda$5007/[O II]$\lambda$3727 versus He II$\lambda$4686/He I$\lambda$5876 diagram to separate AGNs from star-forming galaxies.
[8] See a detailed discussion in the work by Simmonds et al. (2021).

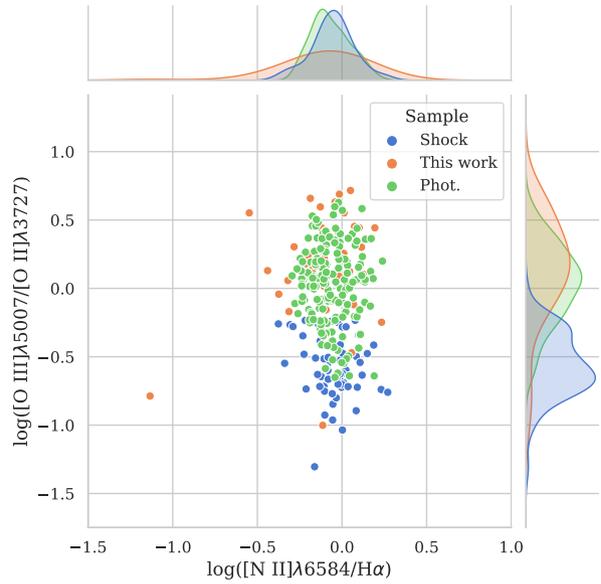

**Figure 3.** Diagram suggested by Dors et al. (2021) to separate shock-dominated and photoionization-dominated objects. Blue and green points represent the results of the models built by Dors et al. (2021) using the SUMA code (Viegas-Aldrovandi & Contini 1989) for shock-dominated and photoionization-dominated objects, respectively. Orange points represent the observational data of our Seyfert 2 sample (see Sect. 2.1). On the top and right sides, we plotted the distributions of the [N II]/H$\alpha$ and [O III]/[O II] lines ratios, respectively, using the same color code.

X-ray Satellite and the Gran Telescopio de Canarias (GTC), found that objects containing high excitation emission lines in their spectra and associated with ultra-luminous X-ray sources have emission lines that lead these objects to occupy the border between star-forming and AGN regions in BTP diagrams. In the top panel of Fig 2, it can be seen that the objects of our sample are located above the maximum starburst line as defined by Kewley et al. (2006). Thus, in principle, we can exclude the ULX ionization from binary stars in the objects of our sample. Finally, concerning fast radiative shocks, as described above we only selected objects with narrow emission lines (FWHM < 1000 km s$^{-1}$), for which low shock contributions to the ionization/heating are expected. It is known the difficulty to identify the presence of shocks based on classical diagnostic diagrams, hence there is an overlap between line intensity ratios predicted by models assuming ionization/heating by photoionization and shocks (see e.g. Kewley et al. 2019; Hirschmann et al. 2023). In this sense, a vast emission lines combination and other nebular parameters (e.g. electron temperatures) have been employed to separate objects accordingly to the AGN ionization source (e.g. Dopita et al. 2016; D'Agostino et al. 2019; Law et al. 2021; Mazzolari et al. 2024). In particular, Dors et al. (2021), by using the SUMA code (Viegas-Aldrovandi & Contini 1989), built detailed composite models assuming photoionization/shock ionization to reproduce narrow optical emission lines of a local sample of Seyfert 2 nuclei. Relying on the results of these composite (AGN+Shock) models, Dors et al. (2021) found that the [O III]$\lambda$5007/[O II]$\lambda$3727 versus [N II]$\lambda$6584/H$\alpha$, [S II]$\lambda$6725/H$\alpha$ and [O I]$\lambda$6300/H$\alpha$ diagrams can be used to separate (with a given uncertainty) shock-dominated and photoionization-dominated objects rather than standard optical diagnostic diagrams. In Fig. 3, one of these diagrams, i.e. [O III]/[O II] versus [N II]/H$\alpha$, containing the SUMA model results from Dors et al. (2021) (blue and green points) and our observational data (orange points) are shown. Also in this figure, the distributions of the emission-line ratios predicted by these two distinct types of models and those for our sample are shown. It can be seen that most of our objects are located in the region occupied by photoionization-dominated models and also present a similar [O III]/[O II] lines ratios distribution to that predicted by such models. Therefore, this result makes it possible to estimate their abundances through the $T_e$-method.

The emission-line intensities (in relation to H$\beta$=1) and the reddening index ($A_v$) for the 38 objects of our sample are listed in Table 1.

## 2.2 Abundance estimates

To derive the nitrogen and oxygen abundances in relation to hydrogen (N/H, O/H), we applied the $T_e$-method adapted for AGNs by Dors et al. (2020b). In this regard, we use the 1.1.13 version of the PYNEB code (Luridiana et al. 2015) to estimate the electron temperature, electron density and ionic abundances (i.e. N$^+$, O$^+$ and O$^{2+}$) for each object in our AGN sample. A detailed description of the $T_e$-method is presented by Dors et al. (2020b) and summarized in what follows.

Initially, the electron temperature representing the high ionization zone ($T_{\rm high}$) and the electron density ($N_e$) were derived, for each object of the sample, from the $RO3$=[O III]($\lambda 4959 + \lambda 5007$)/$\lambda 4363$ and [S II]$\lambda 6716/\lambda 6731$ observational line intensity ratios, respectively. As usual, we assumed [O III]($\lambda 4959/\lambda 5007$)=0.33 (e.g. Storey & Zeippen 2000). For two AGNs the derived temperature values are out of the validity of the $RO3$-$T_{\rm high}$ relation, i.e. their estimated temperature values are higher than $\sim$ 24 000 K (Luridiana et al. 2015). Possibly, for these objects there is a significant contribution by shocks (e.g. Riffel et al. 2021), and hence, they are excluded from our abundance study. The final sample comprehends 38 AGNs with reliable electron temperature and electron density estimates.

For most of the objects of our sample the [N II]$\lambda 5755$ and [O II]$\lambda 7319, \lambda 7330$ auroral emission lines were not measured with a (S/N) > 2, not allowing the direct estimation of the electron temperature for the low ionization zone ($T_{\rm low}$). Thus, for consistency, $T_{\rm low}$ was estimated for all of the objects in our sample through the theoretical relation between $T_{\rm low}$ and $T_{\rm high}$ proposed by Dors et al. (2020b). This relation was obtained through a grid of photoionization models built with the CLOUDY code (Ferland et al. 2013, 2017) simulating Narrow Line Regions (NLRs) of AGNs, and it is given by

$$t_{\rm low} = (a \times t_{\rm high}^3) + (b \times t_{\rm high}^2) + (c \times t_{\rm high}) + d, \quad (2)$$

where a = 0.17, b = −1.07, c = 2.07 and d = −0.33, while $t_{\rm low}$ and $t_{\rm high}$ represent $T_{\rm low}$ and $T_{\rm high}$, respectively, in units of $10^4$ K.

Following the pioneering study by Peimbert & Costero (1969), the ionic abundances were calculated according to:

$$\frac{{\rm O}^{2+}}{{\rm H}^+} = f\left(\frac{[{\rm O\,III}]\lambda 5007}{{\rm H}\beta}, T_{\rm high}, N_e\right), \quad (3)$$

$$\frac{{\rm O}^+}{{\rm H}^+} = f\left(\frac{[{\rm O\,II}]\lambda 3727}{{\rm H}\beta}, T_{\rm low}, N_e\right) \quad (4)$$

and

$$\frac{{\rm N}^+}{{\rm H}^+} = f\left(\frac{[{\rm N\,II}]\lambda 6584}{{\rm H}\beta}, T_{\rm low}, N_e\right). \quad (5)$$

The use of the same electron temperature ($T_{\rm low}$) for the calculations of N$^+$ and O$^+$ is justified by the similarity between the ionization potential (PI) of these ions, i.e. PI(N$^+$)= 29.60 eV and PI(O$^+$)= 35.12 eV (e.g. Garnett 1992; Berg et al. 2021).

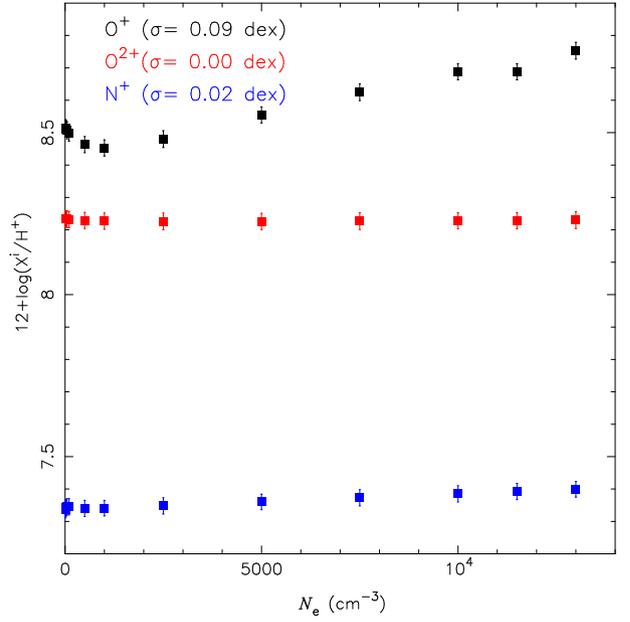

**Figure 4.** Logarithm of ionic abundances for distinct ions (as indicated), derived using the 1.1.13 version of the PYNEB code (Luridiana et al. 2015), as a function of the electron density assumed in the calculations. Observational emission line intensities (see Table 1) of the object spec-6748-56371-0803, used as a benchmark, were used in the simulations. The standard deviation ($\sigma$) values for each ionic abundance are shown. Error bars represent the uncertainty in the estimates.

An obvious concern in the chemical abundance estimates of AGNs is about the electron density that, in general, is higher (by a factor of until $\sim$ 10, e.g. Zhang et al. 2013; Dors et al. 2022; Zhang 2024) than those in SFs (e.g. Copetti et al. 2000), which can produce de-excitation effects in collisionaly lines and, consequently, incorrect ionic abundance values. Spatially resolved observational studies of $N_e$ in NLRs, relying on [S II]$\lambda 6716/\lambda 6731$ line ratio, have shown the existence of density profiles, with $N_e$ ranging from $\sim$ 2500 cm$^{-3}$ in the central parts to $\sim$ 100 cm$^{-3}$ in the outskirts (e.g. Revalski et al. 2018a,b, 2021, 2022; Kakkad et al. 2018; Freitas et al. 2018; Mingozzi et al. 2019; Ruschel-Dutra et al. 2021). Also, $N_e$ estimates in NLRs through the [Ar IV]$\lambda 4711/\lambda 4740$ line ratio, which traces the density in the innermost layers, have found values of up to $\sim$13 000 cm$^{-3}$ (e.g. Vaona et al. 2012; Congiu et al. 2017; Cerqueira-Campos et al. 2021). The lowest critical density ($N_c$) values of the lines involved in the present study are $4.72 \times 10^3$ cm$^{-3}$ ([O II]$\lambda 3716$) and $1.49 \times 10^3$ cm$^{-3}$ ([O II]$\lambda 3731$, see Table 3 of Dors et al. 2023), thus, it is expected some variation of O$^+$ ionic abundances with $N_e$.

Dors et al. (2023) investigated the influence of electron density variation on electron temperature estimates and did not find any significant discrepancy[9]. Here, we analyzed the $N_e$ effects on the oxygen and nitrogen ionic abundances. In this regard, we selected as a benchmark the object of our sample identified as spec-6748-56371-0803, since it presents the lowest derived $N_e$ value (i.e. $\sim$ 20 cm$^{-3}$, see Table 2). For this object, we calculated the ionic abundances by using its emission line intensities (see Table 1) and assuming $N_e$ ranging from 10 to 13 000 cm$^{-3}$. The results of this simulation are shown in Fig. 4, where we can note that the variation of the abundance values with $N_e$ is of the order of 0.16 dex for $O^+$, being the abundance of the other two ions practically constant with $N_e$. Thus, we will consider this value (0.16 dex) as being the uncertainty in $O^+$ abundance estimates due to the possible presence of high density ($N_e > 10\,000$ cm$^{-3}$) clouds in NLRs. However, it is worth mentioning that $O^+$ (PI=35.12 eV) tends to occupy an outer gas region than the one where most of the $Ar^{3+}$ (PI=59.81 eV) is located, therefore, this value must be interpreted as an upper limit.

---

[9] For a more detailed analysis of the effects of $N_e$ on electron temperature estimates see Méndez-Delgado et al. (2023)

**Table 1.** Reddening corrected emission line intensities (in relation to H$\beta$ = 1.0) of the Seyfert 2 sample described in Sect. 2.1. Full table is available in the online material.

| ObjID | [O II] $\lambda$3727 | [O III] $\lambda$4363 | He II 4686 $\lambda$4686 | [O III] $\lambda$5007 | He I $\lambda$5876 | H$\alpha$ $\lambda$6563 | [N II] $\lambda$6584 | [S II] $\lambda$6716 | [S II] $\lambda$6731 | Av |
|---|---|---|---|---|---|---|---|---|---|---|
| spec-6002-56104-0966 | 6.31 ± 0.14 | 0.10 ± 0.02 | 0.18 ± 0.03 | 8.51 ± 0.20 | 0.07 ± 0.04 | 2.86 ± 0.09 | 1.04 ± 0.04 | 0.62 ± 0.03 | 0.56 ± 0.03 | 1.07 |
| spec-5942-56210-0354 | 6.18 ± 0.33 | 0.13 ± 0.02 | 0.18 ± 0.03 | 8.24 ± 0.26 | 0.10 ± 0.04 | 2.86 ± 0.13 | 3.17 ± 0.15 | 1.11 ± 0.06 | 1.02 ± 0.06 | 1.32 |
| spec-7238-56660-0015 | 8.77 ± 0.48 | 0.15 ± 0.05 | 0.15 ± 0.03 | 4.96 ± 0.19 | 0.09 ± 0.04 | 2.86 ± 0.19 | 4.90 ± 0.33 | 1.12 ± 0.09 | 1.02 ± 0.09 | 2.07 |
| spec-3586-55181-0154 | 4.12 ± 0.31 | 0.16 ± 0.03 | 0.18 ± 0.03 | 7.82 ± 0.28 | 0.09 ± 0.04 | 2.86 ± 0.21 | 2.05 ± 0.15 | 0.60 ± 0.06 | 0.61 ± 0.06 | 2.56 |

After the ionic abundance has been derived, we obtain the total abundances (O/H, N/H) following the standard method applied for SFs (e.g. Izotov et al. 2006; Hägele et al. 2008; Berg et al. 2020) and AGNs (e.g. Flury & Moran 2020; Dors et al. 2022). The total abundance of oxygen in relation to the hydrogen (O/H) is obtained assuming the Torres-Peimbert & Peimbert (1977) approach:

$$\frac{O}{H} = \text{ICF}(O^+ + O^{2+}) \times \left[\frac{O^+}{H^+} + \frac{O^{2+}}{H^+}\right], \quad (6)$$

being ICF($O^+ + O^{2+}$) the Ionization Correction Factor (ICF) for oxygen which takes into account the contribution of unobservable oxygen ions. In fact, X-ray (e.g. Cardaci et al. 2009, 2011; Bianchi et al. 2010; Bogdán et al. 2017; Maksym et al. 2019; Kraemer et al. 2020) and infrared (e.g. Diamond-Stanic & Rieke 2012; Fernández-Ontiveros et al. 2016; Fernández-Ontiveros & Muñoz-Darias 2021; Spinoglio et al. 2022) observations of AGNs have shown the presence of emission lines emitted by oxygen ions with higher ionization potential than $O^{2+}$. The ICF($O^+ + O^{2+}$) is derived by assuming (e.g. Izotov et al. 1994; Dors et al. 2022)

$$\text{ICF}(O^+ + O^{2+}) = \frac{He^+ + He^{2+}}{He^+}. \quad (7)$$

The $He^+/H^+$ and $He^{2+}/H^+$ ionic abundance ratios were derived from the observational He I $\lambda$5876/H$\beta$ and He II $\lambda$4696/H$\beta$ line intensity ratios, respectively, measured in the spectrum of each object and following Dors et al. (2022).

For the total nitrogen abundance, we employ the common assumption (Peimbert 1967)

$$\frac{N}{H} = \text{ICF}(N^+) \times \frac{N^+}{H^+}, \quad (8)$$

where

$$\text{ICF}(N^+) = \frac{O}{O^+}. \quad (9)$$

The above empirical ICF relies on the similarity between the ionization potentials of $N^0$ (14.53 eV) and $O^0$ (13.61 eV). Nava et al. (2006), based on photoionization model simulations representing SFs, analyzed the validity of this assumption (i.e. Eq. 9) and found that it is valid at a precision of about 10 per cent. However, the models by Nava and collaborators simulate H II regions with low metallicity [$(Z/Z_\odot) < 0.2$], thus, their result is not necessarily valid for AGNs, for which metallicities near the solar value have been derived (e.g.

Dors et al. 2020a). Moreover, due to the harder ionizing radiation of AGNs in comparison to that of H II regions (e.g. Feltre et al. 2016), somewhat distinct ICF($N^+$) can be derived for these distinct object classes.

To test the applicability of the Eq. 9 for NLRs, we consider the detailed photoionization models built by Dors et al. (2017), by using the Cloudy code (Ferland et al. 2013), to reproduce optical emission lines of 44 local ($z < 0.1$) Seyfert 2 nuclei and derive the N and O nebular abundances. These simulations covered a wide range of metallicity [$(Z/Z_\odot)$=0.2-2.0], number of ionizing photons [$\log Q(H) = 48 - 53$], and electron density [$N_e$=40−900 cm$^{-3}$]. In principle, detailed photoionization models yield more exact predictions of the physical properties of gaseous nebulae than those from grids of photoionization models (e.g. Dors et al. 2014; Morisset et al. 2015; Zhu et al. 2023) because they constrain the space of parameters, i.e. avoid interpretations from unrealistic models. Thus, we take into account the predictions for N/N$^+$ and O/O$^+$ abundance ratios[10] from the 44 detailed models built by Dors et al. (2017). In Fig. 5, a plot of the logarithm of (N/N$^+$) versus (O/O$^+$), obtained using the detailed photoionization model results, are shown. In this figure, the solid line represents the equality between both abundance estimates and the dashed lines their typical uncertainty (40%, e.g. Kennicutt et al. 2003; Hägele et al. 2008).

We can see in Fig. 5 that, for (O/O$^+$) $\lesssim$ 3 (high ionization degree) and within the abundance uncertainty, the photoionization models indicate that Eq. 8 is valid for NLRs. However, for (O/O$^+$) $\gtrsim$ 3 (low ionization degree) the models predict (O/O$^+$) higher than (N/N$^+$). Unfortunately, predictions for the ionization structure (i.e. ionic abundances) of NLRs are rare (or nonexistent) in the literature, making only possible a comparison between the AGN models and H II models. To test if the result above is not an artifact of the NLR models, in Fig. 5 it is also shown:

• Abundance results for H II region photoionization models built by Dors et al. (2018). In these models, the ionizing sources were assumed to be stellar clusters (formed instantaneously) with ages of $10^4$ yr, 2.5 Myr and 6.0 Myr, whose Spectral Energy Distribution was taken from *STARBURST*99 (Leitherer et al. 1999), the logarithm of the ionization parameter ($\log U$) ranging from −4.0 to −1.0, a fixed value of $N_e = 500$ cm$^{-3}$ and $(Z/Z_\odot)$ =0.2-1.0. The reader is referred to the original paper for more details about the photoionization models.

• Results of abundances from the H II region photoionization models by Izotov et al. (2006), built by Stasińska (1990) adopting the PHOTO code (Stasińska 1984) and a single star with distinct values for the effective temperature as ionizing source. These authors proposed expressions for the nitrogen ICF assuming three metallicity

---

[10] The ionic abundance corresponds to that averaged over the volume times the electron density.

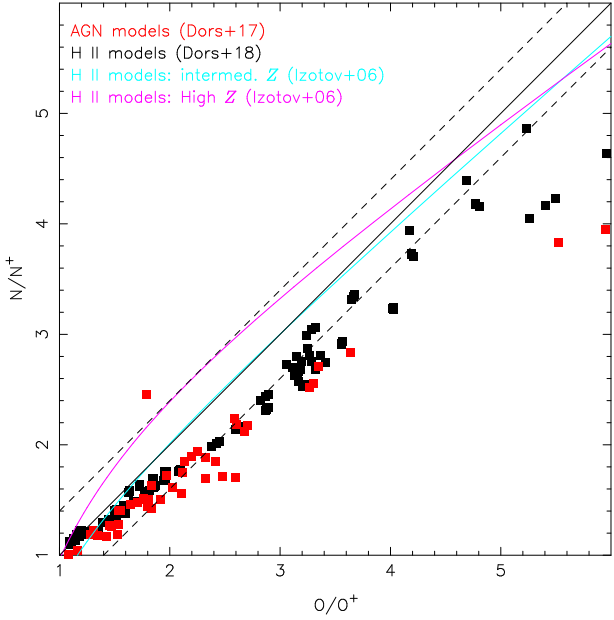

**Figure 5.** $N/N^+$ versus $O/O^+$ abundance ratios. Red points represent the results of detailed photoionization models simulating NLRs built by Dors et al. (2017), by using the CLOUDY code, to reproduce narrow optical emission line ratios of Seyfert 2 nuclei ($z < 0.1$). Black points represent results of H II region models from Dors et al. (2018). Cyan and pink curves represent results of H II region models for the intermediate and high metallicity regime, respectively, (Eqs. 10) proposed by Izotov et al. (2006). Ionic abundances are calculated considering the mean ionization over the gas volume times the electron density. Solid and dashed lines represent the equality between both abundance estimates and their typical errors (40%, e.g. Hägele et al. 2008), respectively.

regimes and given by:

$$\begin{aligned}
\text{ICF}(N^+) &= -0.825v + 0.718 + 0.853/v, \quad \text{low } Z, \\
&= -0.809v + 0.712 + 0.852/v, \quad \text{intermed. } Z, \\
&= -1.476v + 1.752 + 0.688/v, \quad \text{high } Z, \quad (10)
\end{aligned}$$

being $v=O^+/(O^+ + O^{2+})$. We only consider results for the intermediate and high metallicity regimes.

We can see in Fig. 5: (*i*) a similar behavior between NLR model predictions and those from H II region models by Dors et al. (2018). (*ii*) Although the Izotov et al. (2006) models predictions agree with the one-by-one abundance relation, some deviation from them is derived. In any case, the direct abundance results obtained in the present study indicate that most (∼ 90 %) of the objects of our AGN sample present $(O/O^+) \lesssim 3$ (see below), therefore, based on the photoionization model predictions (see Fig. 5), the empirical and usual approach $(N/N^+) = (O/O^+)$ is valid for the NLRs considered in this study.

## 3 RESULTS & DISCUSSION

Since the methodology of the $T_e$-method has been developed by Peimbert & Costero (1969)[11], it has been largely applied for deriving electron temperatures and abundances in thousands of local (e.g.

Smith 1975; Torres-Peimbert et al. 1989; van Zee et al. 1998; Kennicutt et al. 2003; Izotov et al. 2006; Díaz et al. 2007; Hägele et al. 2008; García-Benito et al. 2010; Zurita & Bresolin 2012; López-Hernández et al. 2013; Berg et al. 2020; Rogers et al. 2022) and (very) high redshift (e.g. Jones et al. 2020; Arellano-Córdova et al. 2022; Strom et al. 2023; Sanders et al. 2016, 2020, 2023a; Curti et al. 2023; Laseter et al. 2023; Clarke et al. 2023) SFs. As pointed out above, for AGNs, and in particular for Seyfert 2 nuclei, the $T_e$-method has been applied only for a few (less than 200) local ($z < 0.4$) objects, mainly due to the difficulty to measure auroral lines (see Flury & Moran 2020; Dors et al. 2020a) and also because this reliable method has recently been developed for this object class (Dors et al. 2020b).

Among the heavy elements, nitrogen has prominent optical emission lines ([N II]$\lambda\lambda 6548, 6584$) in AGN spectra (e.g. Osterbrock & Miller 1975; Koski 1978; Richardson et al. 2014; Dopita et al. 2015). For Seyfert 2 nuclei, these prominent [N II] emission lines combined with electron temperature estimations make it possible to obtain reliable estimates of their abundances. The first nitrogen abundance in AGNs by applying the $T_e$-method was carried out by Osterbrock & Miller (1975) for the radio galaxy 3C 405 (Cygnus A), who derived $12+\log(N/H) \sim 8$ which, assuming the solar value $12 + \log(N/H)_\odot = 7.93$ (Holweger 2001), represents $(N/N_\odot) \sim 1.20$. After this pioneering study, Alloin et al. (1992) applied the $T_e$-method for the Seyfert 2 galaxy ESO 138 G1, being only possible to derive the N/O abundance ratio, finding about the solar value $[(N/O) \sim 0.2]$. Finally, Flury & Moran (2020), by using spectroscopic data of NLRs taken from the SDSS DR8 (Aihara et al. 2011) and by using the $T_e$-method, found that AGN nitrogen abundances are typically 1 dex higher than those in H II galaxies with similar oxygen abundances. In the present work, we estimate more precise N/H (and O/H) estimations of NLRs than those in the previous studies since suitable electron temperature values were derived. In fact, for instance, Flury & Moran (2020) estimated $t_{\text{low}}$ through its relation with $t_{\text{high}}$ derived for SFs by Pérez-Montero & Díaz (2003)[12], which is not a good representation for AGNs (Dors et al. 2020b).

As emphasized above, chemical abundance estimates for AGNs through the $T_e$-method, such as for LINERs (e.g. Pérez-Díaz et al. 2021; Oliveira et al. 2024), are rare in the literature, therefore, we compare our results with those for SFs. In view of that, initially, we consider the results for 212 disk H II regions located in five local spiral galaxies (NGC 2403, NGC 628, NGC 5194, NGC 5457, NGC 3184), whose abundances are resulting from the CHAOS project[13] and derived by Berg et al. (2015, 2020), Croxall et al. (2015, 2016) and Rogers et al. (2021). It is worth mentioning that the abundance results of the CHAOS project are in the $8.0 \lesssim 12+\log(O/H) \lesssim 9.0$ metallicity range, in which the nitrogen is dominated by secondary production (Berg et al. 2020). This metallicity range is similar to that derived for Seyfert 2 nuclei (e.g. Dors et al. 2017; Pérez-Montero et al. 2019; Flury & Moran 2020) which make it possible the comparison between the CHAOS and our AGN abundance results. Moreover, the CHAOS abundance results are also derived assuming the $T_e$-method and adopting the same approach for the nitrogen ICF, i.e. $\text{ICF}(N^+)=O/O^+$.

Ionic abundances, ICF values and the resulting total abundances for our Seyfert 2 sample are listed in Table 2.

---

[11] It seems that the first application of the $T_e$-method was carried out by Aller (1954) for the Planetary Nebulae NGC 7027 and by Aller & Liller (1959) for the Orion nebulae.

[12] The $t_{\text{low}}$-$t_{\text{high}}$ relation proposed by Dors et al. (2020b) for AGNs was published after the study of Flury & Moran (2020).
[13] https://www.danielleaberg.com/chaos

**Table 2.** Ionization correction factors and chemical abundances for the Seyfert 2 sample. Full table is available in the online material.

| ID | 12+log($O^+$/H) | 12+log($O^{2+}$/H) | ICF(O) | 12+log(O/H) | 12+log($N^+$/H) | ICF($N^+$) | 12+log(N/H) | log(N/O) |
|---|---|---|---|---|---|---|---|---|
| spec-6002-56104-0966 | 8.58 ± 0.06 | 8.20 ± 0.10 | 1.31 | 8.84 ± 0.08 | 7.44 ± 0.04 | 1.86 | 7.71 ± 0.06 | −1.137 ± 0.112 |
| spec-5942-56210-0354 | 8.52 ± 0.03 | 8.05 ± 0.09 | 1.22 | 8.73 ± 0.05 | 7.89 ± 0.02 | 1.63 | 8.10 ± 0.04 | −0.628 ± 0.069 |
| spec-7238-56660-0015 | 8.73 ± 0.09 | 7.49 ± 0.13 | 1.21 | 8.84 ± 0.09 | 8.11 ± 0.06 | 1.28 | 8.22 ± 0.07 | −0.615 ± 0.124 |
| spec-3586-55181-0154 | 8.35 ± 0.02 | 7.89 ± 0.09 | 1.25 | 8.58 ± 0.04 | 7.69 ± 0.01 | 1.68 | 7.92 ± 0.04 | −0.659 ± 0.057 |

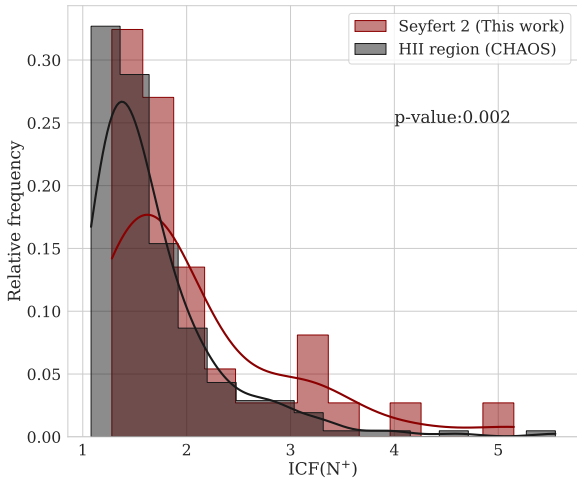

**Figure 6.** Distributions for the ICF($N^+$) = O/$O^+$ derived for our sample of Seyfert 2 nuclei (see Sect. 2) and for disk H ii regions in five local spiral galaxies (NGC 2403, NGC 628, NGC 5194, NGC 5457, NGC 3184) derived by Berg et al. (2015, 2020), Croxall et al. (2015, 2016), and Rogers et al. (2021), as indicated. Range and mean ICF values are listed in Table 3. The p-value from the Anderson-Darling test is indicated.

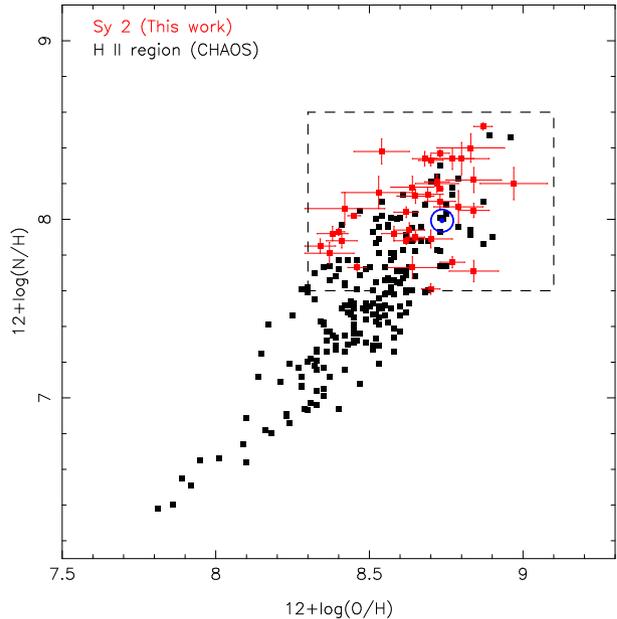

**Figure 7.** Abundance of 12+log(N/H) versus 12+log(O/H). Red points represent abundance estimates for our AGN sample (see Sect. 2). Black points represent disk H ii regions in five local spiral galaxies (NGC 2403, NGC 628, NGC 5194, NGC 5457, NGC 3184), from the CHAOS project, derived by Berg et al. (2015, 2020), Croxall et al. (2015, 2016), and Rogers et al. (2021). Solar symbol represents the nitrogen (12+log(N/H)$_\odot$ = 7.93, Holweger 2001) and oxygen (12+log(O/H)$_\odot$ = 8.69, Allende Prieto et al. 2001) solar values.

### 3.1 Nitrogen ICF

We start our analysis with a comparison between the ICF($N^+$) for our AGN sample and those for disc H ii regions. In Fig. 6, the nitrogen ICF distributions for our AGN sample and for the CHAOS H ii region sample are shown. Both object classes present similar ICF($N^+$) distributions, with the most part (∼ 90%) of objects having values lower than ∼ 3. The p-value from the Anderson-Darling test is 0.002, indicating that both data samples do not follow a similar distribution. However, a larger number of ICF estimates in NLRs is required to produce more reliable statistical conclusions.

In Table 3, the range and the mean ICF values for our sample and CHAOS results are listed. We can see that Seyfert 2 nuclei present a somewhat narrower range of ICF($N^+$) values in comparison to that of H ii regions. However, the mean value for Seyfert 2 nuclei is ∼ 20% higher than for disc H ii regions. That is expected because AGNs tend to have a harder ionizing spectrum than O/B stars (see, for instance, Fig. 2 of Lecroq et al. 2024) resulting in a higher ionization degree than H ii regions. In fact, Feltre et al. (2016), who built a large grid of photoionization models, showed that the intensities of the N v$\lambda$1640/He ii$\lambda$1640 emission line ratio predicted by AGN models are higher (indicating a higher $N^{4+}$ ionic abundance) than those for H ii region models, independently of the metallicity assumed in the models (see also Nakajima & Maiolino 2022).

### 3.2 Nitrogen abundance dispersion

Subsequently, we verify if AGNs and H ii regions follow a similar (N/H)-(O/H) relation and have a distinct scatter of this, as pointed by comparison between observational and predicted emission-line intensity ratios through photoionization models by Dors et al. (2017) and Pérez-Montero et al. (2019). In Fig. 7, the nitrogen [in units of 12+log(N/H)] versus oxygen [in units of 12+log(O/H)] abundance values for the AGN and H ii region samples are shown. The range and mean abundance values for these samples are listed in Table 3. We can note in Fig. 7 that AGN abundances are similar to those of most metallic H ii regions (innermost disk objects), i.e. for the high metallicity regime there is no an extraordinary chemical enrichment of nitrogen in NLRs of AGNs. Interestingly, abundance estimates via the $T_e$-method in NLRs by Dors et al. (2023) indicate that AGNs and SFs present similar S/H abundances for the high-metallicity regime. Probably, as found by Storchi-Bergmann & Pastoriza (1990) from a comparison between photoionization model results and observational data, the abundances of these two elements (N and S) seem to be correlated in AGNs.

To test the scatter in the (N/H)-(O/H) relation, we define the interval of abundance in which AGNs and H ii regions present similar N/H and O/H values. This interval is represented by a box in Fig. 7. Then, we calculated the N/H dispersion for a fixed value of O/H assuming a step of 0.2 dex in the oxygen abundance. We found, for both object classes, a similar dispersion of ∼ 0.2 dex in N/H for fixed O/H values. We pointed out that the same result is derived by taking into account the N/O abundance ratio instead of N/H. Thus, the larger dispersion in the N/O values in comparison to SFs found by Dors et al. (2017) and Pérez-Montero et al. (2019), through photoionization models, seem to be an artifact caused by the (indirect) method used by these authors.

**Table 3.** Range and average nitrogen ICF, N/H, O/H, and N/O relative abundance values for our sample of Seyfert 2s and disk H II regions from the CHAOS project (Berg et al. 2015, 2020; Croxall et al. 2015, 2016; Rogers et al. 2021).

| Object type | ICF(N$^+$) | | 12+log(O/H) | | 12+log(N/H) | | log(N/O) | |
|---|---|---|---|---|---|---|---|---|
| | Range | Average | Range | Average | Range | Average | Range | Average |
| Seyfert 2 | 1.2 - 5.2 | 2.06 | 8.3 - 9.0 | 8.65 ± 0.15 | 7.6 - 8.6 | 8.06 ± 0.22 | −1.2 - −0.2 | −0.58 ± 0.21 |
| H II region | 1.0 - 5.6 | 1.70 | 7.8 - 9.0 | 8.47 ± 0.19 | 6.3 - 8.5 | 7.51 ± 0.38 | −1.5 - −0.4 | −0.96 ± 0.24 |

### 3.3 (N/O)-(O/H) relation

In this section, we discuss our results for log(N/O) versus 12+log(O/H). The (N/O)-(O/H) relation in the ISM of galaxies is paramount to investigate the stellar nucleosynthesis (e.g. Edmunds & Pagel 1978; Alloin et al. 1979) and variations in star-formation rate (e.g. Berg et al. 2020), as well as to constrain chemical evolution models of galaxies (e.g. Henry et al. 2000; Johnson et al. 2023; Dors et al. 2024; Watanabe et al. 2024) and photoionization models (e.g. Kewley & Dopita 2002; Groves et al. 2006; Dors et al. 2014, 2011; Castro et al. 2017; Thomas et al. 2018; Carvalho et al. 2020; Carr et al. 2023).

Over decades, the (N/O)-(O/H) relation has been mainly derived for SFs employing direct estimates of electron temperatures (e.g. Kennicutt et al. 2003; Liang et al. 2006; Hägele et al. 2008, 2012; Dopita et al. 2013; James et al. 2015; Kumari et al. 2018; Berg et al. 2020) and strong-line methods (e.g. Coziol et al. 1999; Pilyugin & Grebel 2016; Belfiore et al. 2017; Schaefer et al. 2020). Concerning AGNs, few previous studies have derived the (N/O)-(O/H) relation for this object class and only (except Flury & Moran 2020) indirect methods have been employed. Carvalho et al. (2020) combined abundances from photoionization modeling by Dors et al. (2017) with estimations for H II regions derived by Pilyugin & Grebel (2016) through the C-method (a strong-line method) and found the relation

$$\log(N/O) = 1.29 \times 12 + \log(O/H) - 11.84 \quad (11)$$

valid for 12 + log(O/H) ≳ 8.0 or, adopting the solar oxygen value 12+log(O/H)$_\odot$=8.69 (Allende Prieto et al. 2001), $(Z/Z_\odot)$ ≳ 0.2. These authors found that Seyfert 2 nuclei obey an N/O versus O/H relation consistent with those of extragalactic H II regions (see also Pérez-Montero et al. 2019).

In the present study, we used a homogeneous dataset and, through the $T_e$-method, we obtained reliable N and O estimates for NLRs of AGNs, allowing us to derive a consistent (N/O)-(O/H) relation. To do that, in Fig. 8, our AGN estimates are combined with those derived also through the $T_e$-method for disk H II regions (CHAOS sample) and for a large sample of H II galaxies, whose direct estimates were taken from the literature. The total number of objects considered in Fig. 8 is 577 (38 Seyfert 2 nuclei, 212 disk H II regions, 327 H II galaxies) spanning in a wide range of metallicity [6.9 < 12 + log(O/H) < 9, or 0.01 < $(Z/Z_\odot)$ < 2.1]. We derive for 12 + log(O/H) > 8.0 [or $(Z/Z_\odot)$ > 0.2] the relation

$$\log(N/O) = [0.90(\pm 0.04) \times 12 + \log(O/H)] - 8.66(\pm 0.40) \quad (12)$$

and for 12 + log(O/H) < 8.0

$$\log(N/O) = -1.42(\pm 0.14). \quad (13)$$

We compare our (N/O)-(O/H) relation with those derived by:

• Dopita et al. (2000): These authors used N and O abundance estimations of disk H II regions by van Zee et al. (1998) and derived the relations

$$\log(N/H) = \begin{cases} -4.57 + \log(Z/Z_\odot); & \log(Z/Z_\odot) \le -0.63 \\ -3.94 + 2 \times \log(Z/Z_\odot); & \text{otherwise.} \end{cases} \quad (14)$$

• Storchi-Bergmann et al. (1994): These authors derived the N and O abundances for a sample of 44 star-forming galaxies and derived the relation

$$\log(N/O) = 0.96 \times 12 + \log(O/H) - 9.29, \quad (15)$$

valid for 12+log(O/H) > 8.4.

We also consider the (N/O)-(O/H) relation given by the Eq. 11 derived by Carvalho et al. (2020). In Fig. 9 the abundance relations above are shown only for the high metallicity regime, i.e. 12+log(O/H) > 8.0. It can be seen that, for fixed O/H values, discrepancies in N/O abundances can reach up to ∼ 0.5 dex if distinct relations are assumed.

### 3.4 Nitrogen-loud AGNs

Baldwin et al. (2003), using observations of the quasi-stellar object (QSO) Q0353-383 ($z = 1.96$) obtained with the Hubble Space Telescope (HST) in the ultraviolet and with the Blanco 4 m telescope at Cerro Tololo Inter-American Observatory (CTIO) in the visible passband, found a new class of QSOs called as nitrogen-loud objects. The main spectral feature of this AGN class is a strong broad N v$\lambda$1240 emission line relative to C iv$\lambda$1549 or to He ii$\lambda$1640, which could be indicative of very high metallicity and nitrogen abundance (∼ 15 times the solar value, see also Batra & Baldwin 2014). Such objects are a rare subclass of AGNs (∼ 1% in SDSS quasars with 1.6 ≲ $z$ ≲ 4, Bentz & Osmer 2004, Bentz et al. 2004, Jiang et al. 2008). Subsequent analysis by Matsuoka et al. (2017), but using narrow optical emission lines, indicate a distinct picture for nitrogen-loud AGNs, i.e. strong nitrogen lines from broad line regions (BLRs) originate from exceptionally high abundances of nitrogen relative to oxygen without very high BLR metallicities (not supersolar metallicities).

Recent JWST/NIRSpec observations of galaxies at $z > 5$ have revealed the presence of galaxies with metallicities around 0.2 $Z_\odot$ and with a very high nitrogen abundance, log(N/O) > −0.5 or (N/O) > 2(N/O)$_\odot$ (e.g. Cameron et al. 2023; Isobe et al. 2023; Topping et al. 2024; Ji et al. 2024). These high-$z$ galaxies can be classified as nitrogen-loud galaxies and have N/O abundance ratio deviating from the locally established (N/O)-(O/H) relation by more than 0.5 dex (Ji et al. 2024). The very high nitrogen abundance in relation to oxygen in these objects has been attributed to the chemical enrichment of the ISM by Wolf Rayet stars (see Ji et al. 2024 and references therein).

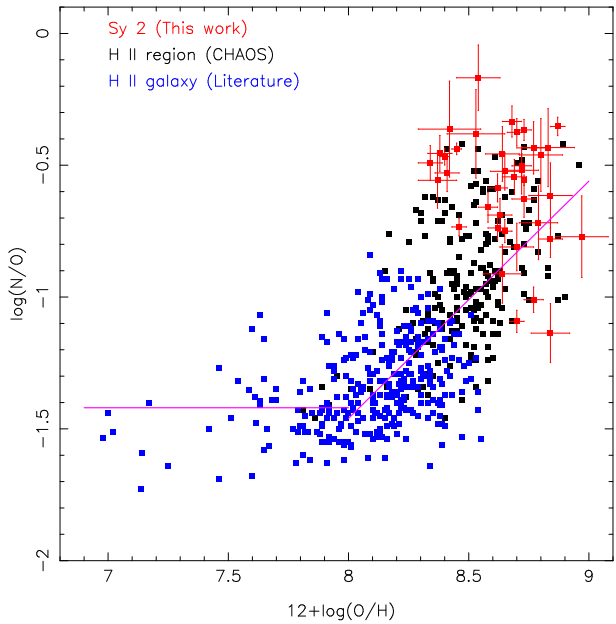

**Figure 8.** log(N/O) versus 12+log(O/H) abundance ratios. Red points represent abundance estimates for our AGN sample (see Sect. 2). Black points represent disk H II regions in five local spiral galaxies (NGC 2403, NGC 628, NGC 5194, NGC 5457, NGC 3184) from the CHAOS project derived by Berg et al. (2015, 2020), Croxall et al. (2015, 2016), and Rogers et al. (2021). Blue points are H II galaxy abundance estimates by Hägele et al. (2006, 2008, 2011, 2012); Skillman et al. (2013); Hirschauer et al. (2016); Izotov et al. (2018); Berg et al. (2019); Kehrig et al. (2020); Thuan et al. (2022); Aver et al. (2022). Lines represent our relations given by Eqs. 12 and 13.

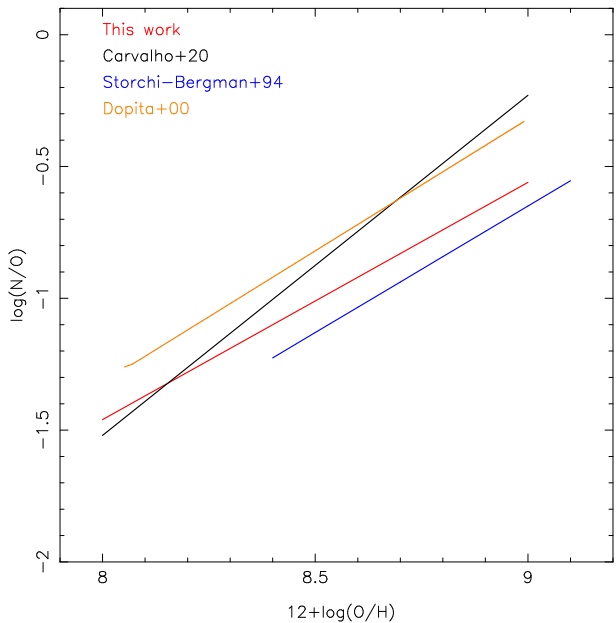

**Figure 9.** Comparison between the (N/O)–(O/H) relations derived in this paper (Eq 12; red line), by Storchi-Bergmann et al. (1994) (Eq. 15; blue line), by Dopita et al. (2000) (Eq. 14; orange line) and by Carvalho et al. (2020) (Eq. 11; black line).

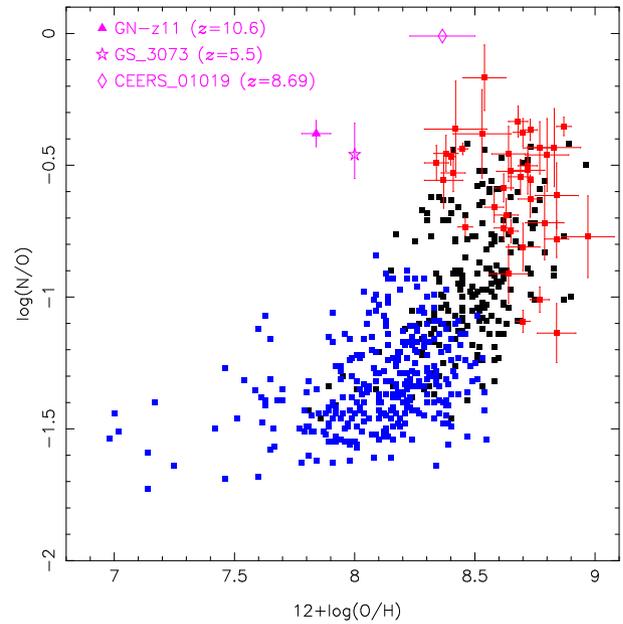

**Figure 10.** As Fig. 8 but showing abundance results for very high-$z$ AGNs derived by Senchyna et al. (2023, triangle), Isobe et al. (2023, diamond) and Ji et al. (2024, star), as indicated.

Our reliable abundance estimates are useful to test if nitrogen-loud AGNs have an excess of N/O abundance in comparison to local active galaxies. To check this, in Fig. 10, a plot of (N/O) versus (O/H), we compare our NLR abundance estimates (as well as those for disk H II regions and H II galaxies) with those for galaxies classified as AGNs at $z > 5$ for which has been derived a nitrogen abundance excess. For that, we consider the objects GN-z11 ($z = 10.6$), GS-3073 ($z = 5.5$) and CEERS-01019 ($z = 8.69$) whose abundance estimates were derived by Senchyna et al. (2023), Ji et al. (2024) and Isobe et al. (2023), respectively. The methods adopted by these authors to estimate their abundances are based on direct detection of auroral lines by using either optical or UV emission lines (Ji et al. 2024) or by comparing observational data to photoionization model results (Senchyna et al. 2023; Isobe et al. 2023). We can see in Fig. 10 that the high-$z$ AGNs show N/O values in consonance with those derived for local NLRs, but have lower O/H abundances. From this comparison, we can make some suppositions:

• The main star-formation event is completed in the early evolution stages of active galaxies, i.e. the epoch of major AGN chemical enrichment has occurred at $z > 5$. This scenario was proposed by Nagao et al. (2006), who did not find evolution of the gas metallicity of the NLR in a QSO sample at redshift $1.5 \lesssim z \lesssim 3.7$ (see also Matsuoka et al. 2009; Dors et al. 2014).

• The bias from the local (N/O)-(O/H) relation can be due to inflows of pristine/low-metallicity gas in the early Universe (e.g. Maiolino et al. 2024), reducing the O/H abundance of high-$z$ galaxies.

Obviously, more direct abundance estimates in AGNs located in a wide range of redshift are necessary to confirm the suppositions above.

## 4 CONCLUSION

In this work, we derive the nitrogen and oxygen abundances in the NLRs of a sample of 38 local ($z < 0.4$) Seyfert 2 nuclei. For that, we consider observational data taken from the SDSS-DR17 and direct estimations of electron temperatures, i.e. adopting the $T_e$-method developed for AGNs. Relying on our results, we conclude:

- Seyfert 2 nuclei present somewhat higher ($\sim 20\%$) ICF values for the $N^+$ ion than disc H II regions. This result is obtained because AGNs tend to have a harder ionizing source than O/B stars, affecting the ionization structure of the nitrogen.
- The dispersion in N/H abundance for a fixed O/H value in AGNs is in agreement with that for disc H II regions with similar metallicity. This finding indicates that both object classes have similar enrichment of nitrogen and oxygen in the interstellar medium.
- Seyfert 2 nuclei follow a similar (N/O)-(O/H) relation than that of star-forming objects. Again, this indicates that these objects have a similar N and O ISM enrichment.
- Active galaxies at $z > 5$ and classified as nitrogen-loud objects show N/O values in consonance with those derived for local NLRs. This is an indication that the major epoch of star formation in host galaxies of AGNs occurred in the early evolutionary stages.

## ACKNOWLEDGEMENTS

OLD is grateful to Fundacão de Amparo à Pesquisa do Estado de São Paulo (FAPESP) and Conselho Nacional de Desenvolvimento Científico e Tecnológico (CNPq). RAR acknowledges the support from CNPq (Proj. 303450/2022-3, 403398/2023-1 & 441722/2023-7) and Fundação Coordenação de Aperfeiçoamento de Pessoal de Nível Superior (CAPES; Proj. 88887.894973/2023-00). CBO is grateful to FAPESP for the support under grant 2023/10182-0. GSI acknowledges financial support from FAPESP (Proj. 2022/11799-9).

## DATA AVAILABILITY

The data underlying this article will be shared on reasonable request to the corresponding author.

## REFERENCES

Abdurro'uf et al., 2022, ApJS, 259, 35
Aihara H., et al., 2011, ApJS, 193, 29
Ali-Dib M., Lin D. N. C., 2023, MNRAS,
Allende Prieto C., Lambert D. L., Asplund M., 2001, ApJ, 556, L63
Aller L. H., 1954, ApJ, 120, 401
Aller L. H., Liller W., 1959, ApJ, 130, 45
Alloin D., Collin-Souffrin S., Joly M., Vigroux L., 1979, A&A, 78, 200
Alloin D., Bica E., Bonatto C., Prugniel P., 1992, A&A, 266, 117
Arellano-Córdova K. Z., et al., 2022, ApJ, 940, L23
Armah M., et al., 2021, MNRAS, 508, 371
Arnett D., 1996, Supernovae and Nucleosynthesis: An Investigation of the History of Matter from the Big Bang to the Present. Princeton University Press
Ascasibar Y., Obreja A. C., Díaz A. I., 2011, MNRAS, 416, 1546
Aver E., Berg D. A., Hirschauer A. S., Olive K. A., Pogge R. W., Rogers N. S. J., Salzer J. J., Skillman E. D., 2022, MNRAS, 510, 373
Baldwin J. A., Phillips M. M., Terlevich R., 1981, PASP, 93, 5
Baldwin J. A., Hamann F., Korista K. T., Ferland G. J., Dietrich M., Warner C., 2003, ApJ, 583, 649
Bär R. E., Weigel A. K., Sartori L. F., Oh K., Koss M., Schawinski K., 2017, MNRAS, 466, 2879
Barlow M. J., Smith L. J., Willis A. J., 1981, MNRAS, 196, 101
Batra N. D., Baldwin J. A., 2014, MNRAS, 439, 771
Belfiore F., et al., 2017, MNRAS, 469, 151
Bentz M. C., Osmer P. S., 2004, AJ, 127, 576
Bentz M. C., Hall P. B., Osmer P. S., 2004, AJ, 128, 561
Berg D. A., Skillman E. D., Garnett D. R., Croxall K. V., Marble A. R., Smith J. D., Gordon K., Kennicutt Robert C. J., 2013, ApJ, 775, 128
Berg D. A., Skillman E. D., Croxall K. V., Pogge R. W., Moustakas J., Johnson-Groh M., 2015, ApJ, 806, 16
Berg D. A., Erb D. K., Henry R. B. C., Skillman E. D., McQuinn K. B. W., 2019, ApJ, 874, 93
Berg D. A., Pogge R. W., Skillman E. D., Croxall K. V., Moustakas J., Rogers N. S. J., Sun J., 2020, ApJ, 893, 96
Berg D. A., Chisholm J., Erb D. K., Skillman E. D., Pogge R. W., Olivier G. M., 2021, ApJ, 922, 170
Bianchi S., Chiaberge M., Evans D. A., Guainazzi M., Baldi R. D., Matt G., Piconcelli E., 2010, MNRAS, 405, 553
Binette L., Krongold Y., Haro-Corzo S. A. R., Humphrey A., Morais S. G., 2023, Rev. Mex. Astron. Astrofis., 59, 113
Bogdán Á., Kraft R. P., Evans D. A., Andrade-Santos F., Forman W. R., 2017, ApJ, 848, 61
Bresolin F., Gieren W., Kudritzki R.-P., Pietrzyński G., Urbaneja M. A., Carraro G., 2009, ApJ, 700, 309
Brum C., Riffel R. A., Storchi-Bergmann T., Robinson A., Schnorr Müller A., Lena D., 2017, MNRAS, 469, 3405
Bykov S. D., Gilfanov M. R., Sunyaev R. A., 2023, arXiv e-prints, p. arXiv:2310.00303
Cameron A. J., Katz H., Rey M. P., Saxena A., 2023, MNRAS, 523, 3516
Cardaci M. V., Santos-Lleó M., Krongold Y., Hägele G. F., Díaz A. I., Rodríguez-Pascual P., 2009, A&A, 505, 541
Cardaci M. V., Santos-Lleó M., Hägele G. F., Krongold Y., Díaz A. I., Rodríguez-Pascual P., 2011, A&A, 530, A125
Cardelli J. A., Clayton G. C., Mathis J. S., 1989, ApJ, 345, 245
Carr D. J., Salzer J. J., Gronwall C., Williams A. L., 2023, ApJ, 955, 141
Carvalho S. P., et al., 2020, MNRAS, 492, 5675
Castro C. S., Dors O. L., Cardaci M. V., Hägele G. F., 2017, MNRAS, 467, 1507
Cerqueira-Campos F. C., Rodríguez-Ardila A., Riffel R., Marinello M., Prieto A., Dahmer-Hahn L. G., 2021, MNRAS, 500, 2666
Cid Fernandes R., Mateus A., Sodré L., Stasińska G., Gomes J. M., 2005, MNRAS, 358, 363
Clarke L., et al., 2023, ApJ, 957, 81
Clayton D. D., 1983, Principles of stellar evolution and nucleosynthesis. University of Chicago Press
Congiu E., et al., 2017, MNRAS, 471, 562
Copetti M. V. F., Mallmann J. A. H., Schmidt A. A., Castañeda H. O., 2000, A&A, 357, 621
Coziol R., Reyes R. E. C., Considère S., Davoust E., Contini T., 1999, A&A, 345, 733
Crowther P. A., 2007, ARA&A, 45, 177
Croxall K. V., Pogge R. W., Berg D. A., Skillman E. D., Moustakas J., 2015, ApJ, 808, 42
Croxall K. V., Pogge R. W., Berg D. A., Skillman E. D., Moustakas J., 2016, ApJ, 830, 4
Curti M., Cresci G., Mannucci F., Marconi A., Maiolino R., Esposito S., 2017, MNRAS, 465, 1384
Curti M., et al., 2023, MNRAS, 518, 425
D'Agostino J. J., et al., 2019, MNRAS, 487, 4153
Daltabuit E., Cox D., 1972, ApJ, 173, L13
Denicoló G., Terlevich R., Terlevich E., 2002, MNRAS, 330, 69
Diamond-Stanic A. M., Rieke G. H., 2012, ApJ, 746, 168
Díaz Tello J., et al., 2017, A&A, 604, A14
Díaz Á. I., Terlevich E., Castellanos M., Hägele G. F., 2007, MNRAS, 382, 251
Dopita M. A., Sutherland R. S., 1996, ApJS, 102, 161
Dopita M. A., Kewley L. J., Heisler C. A., Sutherland R. S., 2000, ApJ, 542, 224
Dopita M. A., Sutherland R. S., Nicholls D. C., Kewley L. J., Vogt F. P. A., 2013, ApJS, 208, 10
Dopita M. A., et al., 2015, ApJS, 217, 12

Dopita M. A., Kewley L. J., Sutherland R. S., Nicholls D. C., 2016, Ap&SS, 361, 61
Dors O. L. J., Krabbe A., Hägele G. F., Pérez-Montero E., 2011, MNRAS, 415, 3616
Dors O. L., Cardaci M. V., Hägele G. F., Krabbe Â. C., 2014, MNRAS, 443, 1291
Dors O. L. J., Arellano-Córdova K. Z., Cardaci M. V., Hägele G. F., 2017, MNRAS, 468, L113
Dors O. L., Agarwal B., Hägele G. F., Cardaci M. V., Rydberg C.-E., Riffel R. A., Oliveira A. S., Krabbe A. C., 2018, MNRAS, 479, 2294
Dors O. L., Monteiro A. F., Cardaci M. V., Hägele G. F., Krabbe A. C., 2019, MNRAS, 486, 5853
Dors O. L., et al., 2020a, MNRAS, 492, 468
Dors O. L., Maiolino R., Cardaci M. V., Hägele G. F., Krabbe A. C., Pérez-Montero E., Armah M., 2020b, MNRAS, 496, 3209
Dors O. L., Contini M., Riffel R. A., Pérez-Montero E., Krabbe A. C., Cardaci M. V., Hägele G. F., 2021, MNRAS, 501, 1370
Dors O. L., et al., 2022, MNRAS, 514, 5506
Dors O. L., et al., 2023, MNRAS, 521, 1969
Dors O. L., Cardaci M. V., Hägele G. F., Ilha G. S., Oliveira C. B., Riffel R. A., Riffel R., Krabbe A. C., 2024, MNRAS, 527, 8193
Edmunds M. G., Pagel B. E. J., 1978, MNRAS, 185, 77P
Ercolano B., Barlow M. J., Storey P. J., Liu X. W., 2003a, MNRAS, 340, 1136
Ercolano B., Morisset C., Barlow M. J., Storey P. J., Liu X. W., 2003b, MNRAS, 340, 1153
Feltre A., Charlot S., Gutkin J., 2016, MNRAS, 456, 3354
Feltre A., et al., 2023, A&A, 675, A74
Ferland G. J., et al., 2013, Rev. Mex. Astron. Astrofis., 49, 137
Ferland G. J., et al., 2017, Rev. Mex. Astron. Astrofis., 53, 385
Fernández-Ontiveros J. A., Muñoz-Darias T., 2021, MNRAS, 504, 5726
Fernández-Ontiveros J. A., Spinoglio L., Pereira-Santaella M., Malkan M. A., Andreani P., Dasyra K. M., 2016, ApJS, 226, 19
Flury S. R., Moran E. C., 2020, MNRAS, 496, 2191
Freitas I. C., et al., 2018, MNRAS, 476, 2760
Galloway M. A., et al., 2015, MNRAS, 448, 3442
García-Benito R., et al., 2010, MNRAS, 408, 2234
Garg P., et al., 2022, ApJ, 926, 80
Garnett D. R., 1992, AJ, 103, 1330
Garnett D. R., Kennicutt Robert C. J., Chu Y.-H., Skillman E. D., 1991, PASP, 103, 850
Gesicki K., Zijlstra A. A., Morisset C., 2016, A&A, 585, A69
Gräfener G., Hamann W. R., 2008, A&A, 482, 945
Groves B. A., Heckman T. M., Kauffmann G., 2006, MNRAS, 371, 1559
Gruenwald R., Viegas S. M., Broguière D., 1997, ApJ, 480, 283
Gusev A. S., Pilyugin L. S., Sakhibov F., Dodonov S. N., Ezhkova O. V., Khramtsova M. S., 2012, MNRAS, 424, 1930
Hägele G. F., Pérez-Montero E., Díaz Á. I., Terlevich E., Terlevich R., 2006, MNRAS, 372, 293
Hägele G. F., Díaz Á. I., Terlevich E., Terlevich R., Pérez-Montero E., Cardaci M. V., 2008, MNRAS, 383, 209
Hägele G. F., García-Benito R., Pérez-Montero E., Díaz Á. I., Cardaci M. V., Firpo V., Terlevich E., Terlevich R., 2011, MNRAS, 414, 272
Hägele G. F., Firpo V., Bosch G., Díaz Á. I., Morrell N., 2012, MNRAS, 422, 3475
Harikane Y., et al., 2023, arXiv e-prints, p. arXiv:2303.11946
Heckman T. M., Best P. N., 2014, ARA&A, 52, 589
Henry R. B. C., Edmunds M. G., Köppen J., 2000, ApJ, 541, 660
Hicks E. K. S., Davies R. I., Maciejewski W., Emsellem E., Malkan M. A., Dumas G., Müller-Sánchez F., Rivers A., 2013, ApJ, 768, 107
Hirschauer A. S., et al., 2016, ApJ, 822, 108
Hirschmann M., et al., 2023, MNRAS, 526, 3610
Holweger H., 2001, in Wimmer-Schweingruber R. F., ed., American Institute of Physics Conference Series Vol. 598, Joint SOHO/ACE workshop "Solar and Galactic Composition". pp 23–30 (arXiv:astro-ph/0107426), doi:10.1063/1.1433974
Huang J., Lin D. N. C., Shields G., 2023, MNRAS, 525, 5702
Hummer D. G., Storey P. J., 1987, MNRAS, 224, 801
Hwang H.-C., et al., 2019, ApJ, 872, 144
Isobe Y., et al., 2023, ApJ, 959, 100
Izotov Y. I., Thuan T. X., 2008, ApJ, 687, 133
Izotov Y. I., Thuan T. X., Lipovetsky V. A., 1994, ApJ, 435, 647
Izotov Y. I., Stasińska G., Meynet G., Guseva N. G., Thuan T. X., 2006, A&A, 448, 955
Izotov Y. I., Thuan T. X., Guseva N. G., Liss S. E., 2018, MNRAS, 473, 1956
James B. L., Koposov S., Stark D. P., Belokurov V., Pettini M., Olszewski E. W., 2015, MNRAS, 448, 2687
Ji X., et al., 2024, arXiv e-prints, p. arXiv:2404.04148
Jiang L., Fan X., Vestergaard M., 2008, ApJ, 679, 962
Jin Y., Kewley L. J., Sutherland R. S., 2022, ApJ, 934, L8
Johnson J. W., Weinberg D. H., Vincenzo F., Bird J. C., Griffith E. J., 2023, MNRAS, 520, 782
Jones T., Sanders R., Roberts-Borsani G., Ellis R. S., Laporte N., Treu T., Harikane Y., 2020, ApJ, 903, 150
Kakkad D., et al., 2018, A&A, 618, A6
Kauffmann G., et al., 2003, MNRAS, 346, 1055
Kauffmann G., et al., 2007, ApJS, 173, 357
Kehrig C., et al., 2011, A&A, 526, A128
Kehrig C., et al., 2020, MNRAS, 498, 1638
Kehrig C., Guerrero M. A., Vílchez J. M., Ramos-Larios G., 2021, ApJ, 908, L54
Kennicutt Robert C. J., Bresolin F., Garnett D. R., 2003, ApJ, 591, 801
Kewley L. J., Dopita M. A., 2002, ApJS, 142, 35
Kewley L. J., Groves B., Kauffmann G., Heckman T., 2006, MNRAS, 372, 961
Kewley L. J., Nicholls D. C., Sutherland R. S., 2019, ARA&A, 57, 511
Kojima T., Ouchi M., Nakajima K., Shibuya T., Harikane Y., Ono Y., 2017, PASJ, 69, 44
Koski A. T., 1978, ApJ, 223, 56
Koss M., et al., 2017, ApJ, 850, 74
Kraemer S. B., Turner T. J., Couto J. D., Crenshaw D. M., Schmitt H. R., Revalski M., Fischer T. C., 2020, MNRAS, 493, 3893
Kudritzki R. P., 2002, ApJ, 577, 389
Kumari N., James B. L., Irwin M. J., Amorín R., Pérez-Montero E., 2018, MNRAS, 476, 3793
LaMassa S. M., Heckman T. M., Ptak A., Urry C. M., 2013, ApJ, 765, L33
Laseter I. H., et al., 2023, arXiv e-prints, p. arXiv:2306.03120
Law D. R., et al., 2021, ApJ, 915, 35
Lecroq M., et al., 2024, MNRAS, 527, 9480
Leitherer C., et al., 1999, ApJS, 123, 3
Liang Y. C., Yin S. Y., Hammer F., Deng L. C., Flores H., Zhang B., 2006, ApJ, 652, 257
López-Hernández J., Terlevich E., Terlevich R., Rosa-González D., Díaz Á., García-Benito R., Vílchez J., Hägele G., 2013, MNRAS, 430, 472
López-Sánchez Á. R., Esteban C., 2008, A&A, 491, 131
Luo Y., et al., 2021, ApJ, 908, 183
Luridiana V., Morisset C., Shaw R. A., 2015, A&A, 573, A42
Maiolino R., et al., 2023, arXiv e-prints, p. arXiv:2308.01230
Maiolino R., et al., 2024, Nature, 627, 59
Maksym W. P., et al., 2019, ApJ, 872, 94
Mallmann N. D., et al., 2018, MNRAS, 478, 5491
Marino R. A., et al., 2013, A&A, 559, A114
Matsuoka K., Nagao T., Maiolino R., Marconi A., Taniguchi Y., 2009, A&A, 503, 721
Matsuoka K., Nagao T., Maiolino R., Marconi A., Park D., Taniguchi Y., 2017, A&A, 608, A90
Matsuoka K., Nagao T., Marconi A., Maiolino R., Mannucci F., Cresci G., Terao K., Ikeda H., 2018, A&A, 616, L4
Mayya Y. D., et al., 2020, MNRAS, 498, 1496
Mazzalay X., Rodríguez-Ardila A., Komossa S., 2010, MNRAS, 405, 1315
Mazzolari G., et al., 2024, arXiv e-prints, p. arXiv:2404.10811
Méndez-Delgado J. E., et al., 2023, MNRAS, 523, 2952
Meynet G., Maeder A., 2002, A&A, 381, L25
Mingozzi M., et al., 2019, A&A, 622, A146
Molina M., Reines A. E., Latimer L. J., Baldassare V., Salehirad S., 2021, ApJ, 922, 155

Monteiro A. F., Dors O. L., 2021, MNRAS, 508, 3023
Monteiro H., Schwarz H. E., Gruenwald R., Heathcote S., 2004, ApJ, 609, 194
Morisset C., Stasińska G., Peña M., 2005, MNRAS, 360, 499
Morisset C., Delgado-Inglada G., Flores-Fajardo N., 2015, Rev. Mex. Astron. Astrofis., 51, 103
Nagao T., Maiolino R., Marconi A., 2006, A&A, 447, 863
Nakajima K., Maiolino R., 2022, MNRAS, 513, 5134
Nava A., Casebeer D., Henry R. B. C., Jevremovic D., 2006, ApJ, 645, 1076
Nazé Y., Rauw G., Manfroid J., Chu Y. H., Vreux J. M., 2003, A&A, 408, 171
Nicholls D. C., Sutherland R. S., Dopita M. A., Kewley L. J., Groves B. A., 2017, MNRAS, 466, 4403
Nomoto K., Kobayashi C., Tominaga N., 2013, ARA&A, 51, 457
Norman C., Scoville N., 1988, ApJ, 332, 124
Oh K., et al., 2022, ApJS, 261, 4
Oliveira C. B. J., Krabbe A. C., Dors O. L. J., Zinchenko I. A., Hernandez-Jimenez J. A., Cardaci M. V., Hägele G. F., Ilha G. S., 2024, arXiv e-prints, p. arXiv:2404.16631
Osterbrock D. E., Miller J. S., 1975, ApJ, 197, 535
Pagel B. E. J., Edmunds M. G., 1981, ARA&A, 19, 77
Pagel B. E. J., Edmunds M. G., Smith G., 1980, MNRAS, 193, 219
Peimbert M., 1967, ApJ, 150, 825
Peimbert M., Costero R., 1969, Boletin de los Observatorios Tonantzintla y Tacubaya, 5, 3
Peimbert M., Torres-Peimbert S., Rayo J. F., 1978, ApJ, 220, 516
Peimbert M., Peimbert A., Delgado-Inglada G., 2017, PASP, 129, 082001
Pérez-Díaz B., Masegosa J., Márquez I., Pérez-Montero E., 2021, MNRAS, 505, 4289
Pérez-Montero E., 2017, PASP, 129, 043001
Pérez-Montero E., Contini T., 2009, MNRAS, 398, 949
Pérez-Montero E., Díaz A. I., 2003, MNRAS, 346, 105
Pérez-Montero E., et al., 2013, A&A, 549, A25
Pérez-Montero E., Dors O. L., Vílchez J. M., García-Benito R., Cardaci M. V., Hägele G. F., 2019, MNRAS, 489, 2652
Pérez-Montero E., Kehrig C., Vílchez J. M., García-Benito R., Duarte Puertas S., Iglesias-Páramo J., 2020, A&A, 643, A80
Pilyugin L. S., 1999, A&A, 346, 428
Pilyugin L. S., Grebel E. K., 2016, MNRAS, 457, 3678
Plat A., Charlot S., Bruzual G., Feltre A., Vidal-García A., Morisset C., Chevallard J., Todt H., 2019, MNRAS, 490, 978
Prieto M. A., Fernandez-Ontiveros J. A., Bruzual G., Burkert A., Schartmann M., Charlot S., 2019, MNRAS, 485, 3264
Ramón-Fox F. G., Aceves H., 2020, MNRAS, 491, 3908
Rayo J. F., Peimbert M., Torres-Peimbert S., 1982, ApJ, 255, 1
Rembold S. B., et al., 2017, MNRAS, 472, 4382
Revalski M., Crenshaw D. M., Kraemer S. B., Fischer T. C., Schmitt H. R., Machuca C., 2018a, ApJ, 856, 46
Revalski M., et al., 2018b, ApJ, 867, 88
Revalski M., et al., 2021, ApJ, 910, 139
Revalski M., et al., 2022, ApJ, 930, 14
Ricci C., et al., 2017, MNRAS, 468, 1273
Richardson C. T., Allen J. T., Baldwin J. A., Hewett P. C., Ferland G. J., 2014, MNRAS, 437, 2376
Rickards Vaught R. J., Sandstrom K. M., Hunt L. K., 2021, ApJ, 911, L17
Riffel R. A., et al., 2016, MNRAS, 461, 4192
Riffel R. A., et al., 2021, MNRAS, 501, L54
Riffel R., et al., 2023, MNRAS, 524, 5640
Rogers N. S. J., Skillman E. D., Pogge R. W., Berg D. A., Moustakas J., Croxall K. V., Sun J., 2021, ApJ, 915, 21
Rogers N. S. J., Skillman E. D., Pogge R. W., Berg D. A., Croxall K. V., Bartlett J., Arellano-Córdova K. Z., Moustakas J., 2022, ApJ, 939, 44
Ruschel-Dutra D., Dall'Agnol De Oliveira B., 2020, danielrd6/ifscube v1.0, Zenodo, doi:10.5281/zenodo.3945237
Ruschel-Dutra D., et al., 2021, MNRAS, 507, 74
Sanders R. L., et al., 2016, ApJ, 825, L23
Sanders R. L., et al., 2020, MNRAS, 491, 1427
Sanders R. L., Shapley A. E., Topping M. W., Reddy N. A., Brammer G. B., 2023a, arXiv e-prints, p. arXiv:2303.08149
Sanders R. L., Shapley A. E., Topping M. W., Reddy N. A., Brammer G. B., 2023b, ApJ, 955, 54
Sanders R. L., Shapley A. E., Topping M. W., Reddy N. A., Brammer G. B., 2024, ApJ, 962, 24
Sartori L. F., Schawinski K., Treister E., Trakhtenbrot B., Koss M., Shirazi M., Oh K., 2015, MNRAS, 454, 3722
Schaefer A. L., Tremonti C., Belfiore F., Pace Z., Bershady M. A., Andrews B. H., Drory N., 2020, ApJ, 890, L3
Schaefer A. L., et al., 2022, ApJ, 930, 160
Schaerer D., Fragos T., Izotov Y. I., 2019, A&A, 622, L10
Scholtz J., et al., 2023, arXiv e-prints, p. arXiv:2311.18731
Schönell A. J., Storchi-Bergmann T., Riffel R. A., Riffel R., Bianchin M., Dahmer-Hahn L. G., Diniz M. R., Dametto N. Z., 2019, MNRAS, 485, 2054
Searle L., 1971, ApJ, 168, 327
Senchyna P., Plat A., Stark D. P., Rudie G. C., 2023, arXiv e-prints, p. arXiv:2303.04179
Shields G. A., Searle L., 1978, ApJ, 222, 821
Shirazi M., Brinchmann J., 2012, MNRAS, 421, 1043
Sidoli F., Smith L. J., Crowther P. A., 2006, MNRAS, 370, 799
Simmonds C., Schaerer D., Verhamme A., 2021, A&A, 656, A127
Skillman E. D., et al., 2013, AJ, 146, 3
Smith H. E., 1975, ApJ, 199, 591
Spinoglio L., Fernández-Ontiveros J. A., Malkan M. A., 2022, ApJ, 941, 46
Stasińska G., 1984, A&AS, 55, 15
Stasińska G., 1990, A&AS, 83, 501
Storchi-Bergmann T., Pastoriza M. G., 1990, PASP, 102, 1359
Storchi-Bergmann T., Calzetti D., Kinney A. L., 1994, ApJ, 429, 572
Storchi-Bergmann T., Raimann D., Bica E. L. D., Fraquelli H. A., 2000, ApJ, 544, 747
Storchi-Bergmann T., González Delgado R. M., Schmitt H. R., Cid Fernandes R., Heckman T., 2001, ApJ, 559, 147
Storchi-Bergmann T., Dors Oli L. J., Riffel R. A., Fathi K., Axon D. J., Robinson A., Marconi A., Östlin G., 2007, ApJ, 670, 959
Storey P. J., Zeippen C. J., 2000, MNRAS, 312, 813
Strom A. L., et al., 2023, ApJ, 958, L11
Talbot R. J. J., Arnett D. W., 1974, ApJ, 190, 605
Thomas A. D., Dopita M. A., Kewley L. J., Groves B. A., Sutherland R. S., Hopkins A. M., Blanc G. A., 2018, ApJ, 856, 89
Thuan T. X., Izotov Y. I., 2005, ApJS, 161, 240
Thuan T. X., Guseva N. G., Izotov Y. I., 2022, MNRAS, 516, L81
Topping M. W., et al., 2024, MNRAS, 529, 3301
Torres-Peimbert S., Peimbert M., 1977, Rev. Mex. Astron. Astrofis., 2, 181
Torres-Peimbert S., Peimbert M., Fierro J., 1989, ApJ, 345, 186
Tozzi G., Maiolino R., Cresci G., Piotrowska J. M., Belfiore F., Curti M., Mannucci F., Marconi A., 2023, MNRAS, 521, 1264
Übler H., et al., 2023, A&A, 677, A145
Vaona L., Ciroi S., Di Mille F., Cracco V., La Mura G., Rafanelli P., 2012, MNRAS, 427, 1266
Viegas-Aldrovandi S. M., Contini M., 1989, ApJ, 339, 689
Vilchez J. M., Esteban C., 1996, MNRAS, 280, 720
Vincenzo F., Kobayashi C., 2018, MNRAS, 478, 155
Wang Q., Kron R. G., 2020, MNRAS, 498, 4550
Wang Y., Lin D. N. C., Zhang B., Zhu Z., 2023, arXiv e-prints, p. arXiv:2310.00038
Watanabe K., et al., 2024, ApJ, 962, 50
Wood K., Mathis J. S., Ercolano B., 2004, MNRAS, 348, 1337
Woosley S. E., Weaver T. A., 1995, ApJS, 101, 181
York D. G., et al., 2000, AJ, 120, 1579
Zaragoza-Cardiel J., Gómez-González V. M. A., Mayya D., Ramos-Larios G., 2022, MNRAS, 514, 1689
Zhang X., 2024, ApJ, 960, 108
Zhang Z. T., Liang Y. C., Hammer F., 2013, MNRAS, 430, 2605
Zhu P., Kewley L. J., Sutherland R. S., 2023, ApJ, 954, 175
Zurita A., Bresolin F., 2012, MNRAS, 427, 1463
do Nascimento J. C., et al., 2022, MNRAS, 513, 807

van Zee L., Salzer J. J., Haynes M. P., O'Donoghue A. A., Balonek T. J., 1998, AJ, 116, 2805

This paper has been typeset from a T<sub>E</sub>X/L<sup>A</sup>T<sub>E</sub>X file prepared by the author.